%% file: FromOracleToTM.tex
\documentclass{jicspack}
\input{config}

\usepackage{enumerate}
\usepackage{graphicx}

\usepackage{amssymb}

\begin{document}

\begin{premaker}


\title{Inquiry of \textit{P-reduction} in Cook's 1971 Paper\\
 - from  \textit{Oracle machine} to \textit{Turing machine}}
\author{JianMing ZHOU, Yu LI}
\ead{yu.li@u-picardie.fr}
\address{(1) MIS, Universit\'{e} de Picardie Jules Verne, 33 rue Saint-Leu, 80090 Amiens, France \\
(2) Institut of Computational Theory and Application, Huazhong University of Science and Technology, Wuhan, China}


\begin{abstract}

In this paper, we inquire the key concept \textit{P-reduction} in \textit{Cook's theorem} and reveal that there exists the fallacy of definition in \textit{P-reduction} caused by the disguised displacement of \textit{NDTM} from \textit{Oracle machine} to \textit{Turing machine}. The definition or derivation of  \textit{P-reduction} is essentially equivalent to Turing's  \textit{computability}. Whether \textit{NP} problems might been reduced to logical forms (\textit{tautology} or \textit{SAT}) or \textit{NP} problems might been reduced each other, they have not been  really proven in Cook's 1971 paper. 

\end{abstract}

\end{premaker}

\section{Introduction}

\textit{NP} problems and computational complexity theory are  based on \textit{Cook's theorem}  proposed by Cook in the paper entitled  \textit{The Complexity of Theorem-Proving Procedures} in 1971 \cite{cook1}.  

\textit{Cook's theorem} claims that any problem decidable by  \textit{NDTM (Non-Deterministic Turing Machine)} can be reduced to the \textit{SAT (SATisfiability)} problem in polynomial time by  \textit{TM (Turing Machine)} .

From \textit{Cook's theorem}, on the one hand, it produces out two popular definitions of \textit{NP (Non-deterministic Polynomial time)} that are considered to be equivalent: 

\textbf{Definition 1 }
\textit{NP is the class of problems decidable (or solvable) by NDTM in polynomial time.}
 
 \textbf{Definition 2 }
\textit{NP is the class of problems verifiable by TM in polynomial time. }

On the other hand, it deduces out the \textit{NP-completeness}s: if there is a polynomial time algorithm to solve the  \textit{SAT} problem, then each problem in \textit{NP}  can be solved by a polynomial time algorithm. The question of whether such a polynomial time algorithm exists for solving a \textit{NP}  problem is expressed as the \textit{P vs NP}  problem that is widely considered as the most important open problem in computer science, also selected as one of the seven millennial challenges by the Clay Mathematics Institute in 2001  \cite{cook2}.

However, people who come across the theory of \textit{NP-completeness}  would have more or less such impressions, although the definition of \textit{NP} seems simple and clear, but if one slightly investigates it, one would feel some confusion that is difficult to express out \cite{stackexchange}. Scott Aaroson who perennially works on \textit{P vs NP}  expressed his perplexity:\textit{There seems to be an "invisible electric fence" that separates the P problem from the NP complete problem} \cite{scott1}\cite{scott2}. Moreover, if one wants to go deeper into \textit{NP} completeness theory, one would have to go back to study or think this theory again and again,  finally one is either disappointed or obligated himself to obey the authoritative explications. On the other hand, \textit{NP} completeness theory is almost completely out of touch with the practical resolution of \textit{NP} problems, some people even say that this theory can be abandoned. In fact, the content of \textit{NP} completeness theory is getting lighter and lighter in basic computer theory courses in universities. These situations prevent computer theory from  further developing. In fact, the \textit{P vs NP} problem has implicitly become the gap between the computer theory and the booming artificial intelligence theory.

We have made some preliminary interpretations of several issues related to \textit{Cook's theorem} from different perspectives \cite{art1}\cite{art2}\cite{art3}\cite{art4}. In this paper,  we inquire the key concept \textit{P-reduction} of \textit{Cook's theorem} and reveal the cognitive and theoretical errors in \textit{Cook's theorem}. We hope to attract the attention of the academic community.

\section{Overview of \textit{Cook's theorem}} 

The original statement of \textit{Cook's theorem} was presented in Cook's 1971 paper  \cite{cook1}:

\textbf{Theorem 1}
\textit{If a set $S$ of strings is accepted by some nondeterministic Turing machine within polynomial time, then $S$ is $P$-reducible to \{$DNF$ tautologies\}}.

The main idea of the proof of Theorem 1 was described in \cite{cook1}: 

 \textit{Proof of the theorem. Suppose a nondeterministic Turing machine $M$ accepts a set $S$ of strings within time $Q(n)$, where $Q(n)$ is a polynomial. Given an input $w$ for $M$, we will construct a propositional formula $A(w)$ in conjunctive normal form ($CNF$) such that $A(w)$ is satisfiable iff $M$ accepts $w$.  Thus $\neg A(w)$ is easily put in disjunctive normal form (using De MorganÕs laws), and $\neg A(w)$ is a tautology if and only if w $ \not \in S$. Since the whole construction can be carried out in time bounded by a polynomial in $\mid w \mid$ (the length of $w$), the theorem will be proved.}\\

As an intuitive interpretation, Cook attempted to prove that any problem solvable by a  \textit{nondeterministic Turing machine} at level of cognition is \textit{P-reducible}  to the problem of determining whether a given propositional formula is satisfiable at level  of computation.

 \section{\textit{P-reduction} and \textit{Turing reduction} : disguised displacement of \textit{NDTM}}

 \textit{P-reduction} is defined as follows  \cite{cook1}:
 
 \textbf{Definition.}
\textit{A set S of strings is P-reducible (P for polynomial) to a set T of strings iff there is some query machine M and a polynomial Q(n) such that for each input string w, the T-computation of M with input w halts within $Q(\mid w \mid)$ steps ($\mid w \mid$ is the length of w) and ends in an accepting state iff $w \in S$.}

About the \textit{query machine}, Cook said \cite{cook1}:

  \textit{In order to make this notion precise, we introduce query machines, which are like Turing machines with oracles in  \cite{cook1}.} \\
  
 \textit{A query machine is a multitape Turing machine with a distinguished tape called the query tape, and three distinguished states called the $query~state$, $yes~state$, and $no~state$, respectively. If $M$ is a query machine and $T$ is a set of strings, then a $T$-computation of $M$ is a computation of $M$ in which initially $M$ is in the initial state and has an input string $w$ on its input tape,  and each time $M$ assures the query state there is a string $u$ on the query tape,  and the next state $M$ assumes is the yes state if $u \in  T$ and the no state if $u  \not  \in T$.  We think of an 'oracle', which knows $T$, placing $M$ in the yes state or no state}.  \\
  
It can be seen that the \textit{NDTM} in  \textit{Cook's theorem} refers to the  \textit{query machine}, and the  \textit{query machine} refers to the \textit{Oracle Machine}. Thus,  \textit{NP} is decidable by \textit{NDTM}, actually that is to say, \textit{NP} is decidable by the \textit{Oracle  Machine} (\textit{Definition 1}), and \textit{P-reduction} refers to \textit{Turing reduction} based on \textit{Oracle  Machine} (Figure 1).

\begin{figure} [h]
\begin{center}
\includegraphics[scale=.3]{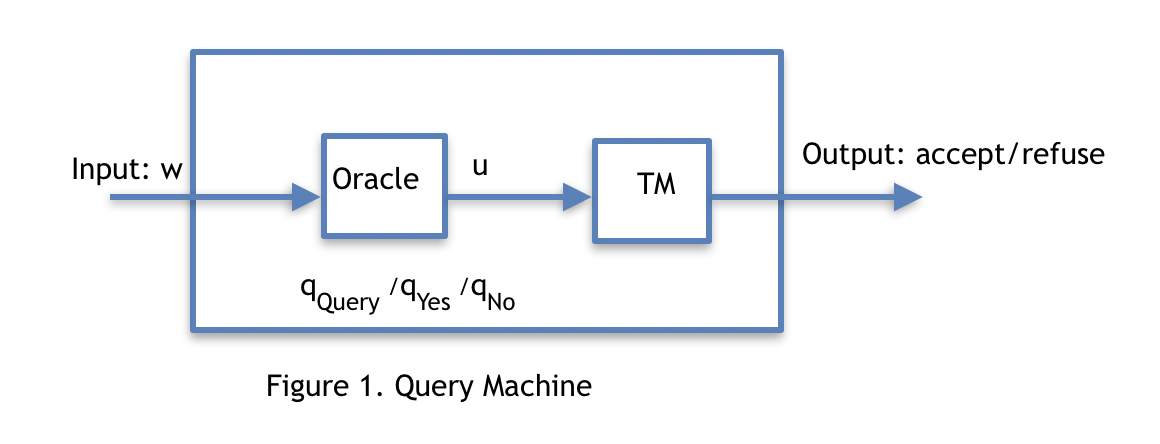}
\end{center}
\label{fig1}
\end{figure}

Concerning \textit{Turing reduction}, Martin Davis said in his paper entitled \textit{what is Turing Reducibility?} \cite{Martin}:

 \textit{The concept of Turing reducibility has to do with the question: can one non-computable set be more non-computable than another? In a rather incidental aside to the main topic of Alan Turing’s doctoral dissertation (the subject of Soloman Feferman’s article in this issue of the Notices), he introduced the idea of a computation with respect to an oracle. An oracle for a particular set of natural numbers may be visualized as a ‘black box’ that will correctly answer questions about whether specific numbers belong to that set. We can then imagine an oracle algorithm whose operations can be interrupted to query such an oracle with its further progress dependent on the reply obtained. Then for sets A,B of natural numbers, A is said to be Turing Reducible to B if there is an oracle algorithm for testing membership in A having full recourse to an oracle for B. The notation used is: $A \leqslant t B$. Of course, if B is itself a computable set, then nothing new happens; in such a case $A \leqslant t B$ just means that A is computable. But if B is non-computable, then interesting things happen}.  \\ 

According to Martin Davis's understanding,  \textit{Turing reduction} can be defined as follows: a set $A$ (in number theory) is reduced to a set $B$ (in number theory), if $B$ can be decided by some oracle (black box), then $A$ can be decided. However, some sets in number theory themself are undecidable, and only an oracle can decide them, thus Martin Davis said, \textit{if B is itself a computable set, then nothing new happens}, that is, \textit{Turing reduction}  is unconsciously considered to deal with uncomputable problems.

Therefore, \textit{Oracle} should be understood as an omniscient logic or an omnipotent algorithm. In this sense, \textit{Oracle  Machine} is not a constructive model of machine, but an imaginary theoretical model, as Turing said \cite{Turing},  \textit{We shall not go any further into the nature of this oracle apart from saying that it cannot be a machine}. In other words, \textit{Oracle  Machine}  is   \textit{NTM (Non Turing Machine)} as opposed to \textit{TM (Turing Machine)}.

From the perspective of the relativity to computability, the introduction of \textit{NP} by  \textit{Oracle  Machine} to study its \textit{undecidablility} is meaningful. Unfortunately, Cook completely misunderstood Turing's original idea about \textit{Oracle  Machine}, and tried to use \textit{Oracle  Machine} as an actual machine in his constructive proof. However, the \textit{Oracle  Machine} is just a theoretical model, so it is impossible to actually represent the solution to a problem given by \textit{Oracle}, that is, the string $u$ (Figure 1) in the \textit{query machine}, so Cook had to secretly throw away \textit{Oracle} before he began his proof:

  \textit{We may as well assume the Turing machine M has only one tape, which is infinite to the right but has a left-most square. Let us number the squares from left to right 1,2,.... L et us fix an input w to M of length n, and suppose $w \in S$. Then there is a computation of M with input w that ends in an accepting state within T = Q(n) steps. }

In other words, Cook used an \textit{invisible hand}   to directly incorporate $u$ obtained by \textit{Oracle} from the query tape onto the tape of \textit{Turing Machine} without any justification, so that the \textit{query machine} is directly transformed into  \textit{Turing Machine} from \textit{Oracle  Machine} (Figure 2):

\begin{figure} [h]
\begin{center}
\includegraphics[scale=.3]{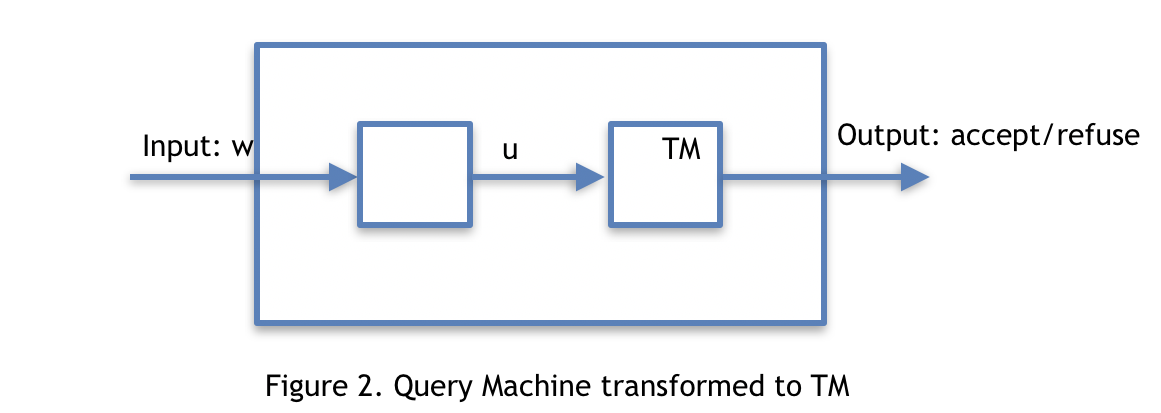}
\end{center}
\label{fig2}
\end{figure}

Later scholars maybe thought that such disguised displacement of \textit{NDTM} was not appropriate. In order to save \textit{Cook's theorem}, a guessing module was used to replace the  \textit{Oracle}  in the \textit{query machine}  (Figure 3) \cite{Karp}\cite{garey}:

\begin{figure} [h]
\begin{center}
\includegraphics[scale=.3]{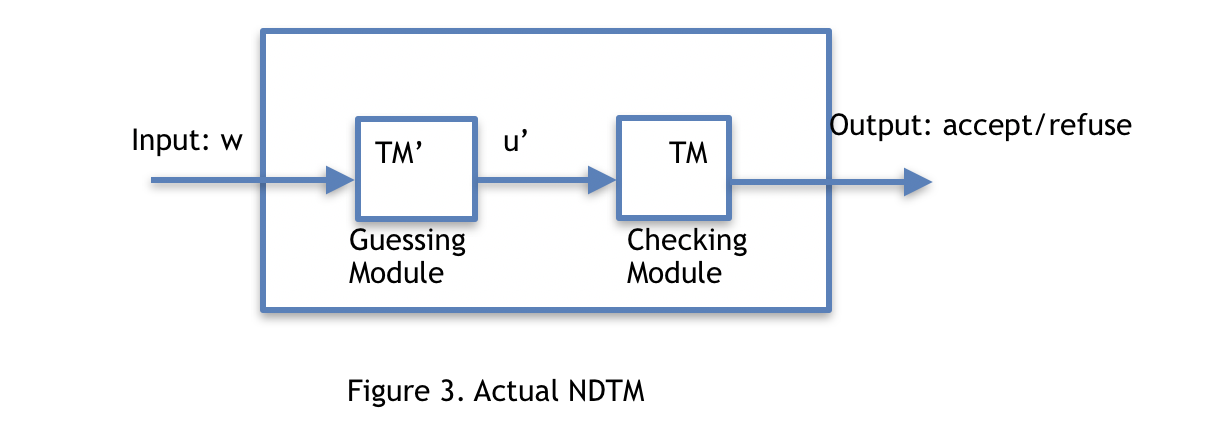}
\end{center}
\label{fig3}
\end{figure}

This \textit{NDTM} with guessing module is just the current \textit{NDTM} \cite{sipser}:

 \textit{At any point in a computation the machine may proceed according to several possibilities. The computation of a nondeterministic Turing machine is a tree whose branches correspond to different possibilities for the machine. If some branch of the computation leads to the accept state, the machine accepts its input.}

And this \textit{NDTM} is in essence  \textit{TM} \cite{sipser}: 

 \textit{Every nondeterministic Turing machine has an equivalent deterministic Turing machine. (Theorem 3.16 in [9]) }

In this way, the solution $u$ to a problem $w$ that should been given by the \textit{Oracle}  of the query machine becomes a guess solution $u’$ (certificate) given by the guessing module \textit{$TM'$} and then $u’$ is verified by an other Turing machine \textit{TM}, which leads to the second popular definition of \textit{NP}:  \textit{NP is polynomial time verifiable (Definition 2)} .

So far, \textit{Cook's theorem} completed the disguised displacement of \textit{NDTM} from \textit{Oracle machine}  to \textit{Turing Machine},  and caused the confusion between the two mutually exclusive concepts \textit{NTM} and \textit{TM}! Henceforth, the undecidable problems (\textit{NP}) expressed by \textit{Oracle Machine} became decidable problems (\textit{P}) expressed  by \textit{Turing Machine}, which leads to the  \textit{loss of nondeterminism from NP}. 

 \textit{P-reduction}  can no longer refer to  \textit{Turing reduction}, then \textit{Cook's theorem}  has lost any support in mathematical and logical sense.

\section{Conclusion}

Our analysis shows  that there exists the fallacy of definition in    \textit{P-reduction} of \textit{Cook's theorem}, caused by the disguised displacement of \textit{NDTM} from \textit{Oracle machine} to \textit{Turing machine}. Conseqently  \textit{Cook's theorem} declares that \textit{NP} problems can be reduced  to logical forms (\textit{tautology} or \textit{SAT}) through \textit{TM}, which is equivalent to affirming the \textit{computability} of  \textit{NP} problems, thus  the \textit{loss of nondeterminism from NP}.  A brief analysis about the logical form of $A(w)$,  a key of the proof of \textit{Cook's theorem}, can be found in \cite{art5}.

The impact of \textit{Cook's theorem} is enormous, it concerns \textit{NP}, \textit{NP-completeness}, and the relationship between computability theory and  computational complexity theory, so that \textit{P vs NP} has become one of millennial challenges. This issue is not only an academic theoretical one, but also  a historical event with great influence in modern academic history.

In introducing the second poll about  \textit{P vs NP}  conducted by Gasarch in 2012, Hemaspaandra said \cite{william}:

\textit{I hope that people in the distant future will look at these four articles to help get a sense of people’s thoughts back in the dark ages when P versus NP had not yet been resolved.}


\end{document}

%% file: config.tex
\setvolume{8}                             
\setyear{2012}                             
\setpagerange{1}{8}                    
\setheadauthor{X. Liu et al.}          
\setissn{1553--9105} \setpubdate{January 2012} \setno{1}

\afterpage{\beginheader}                   